\documentclass[aps,preprint,groupedaddress,floatfix]{revtex4}
\usepackage{graphics}
\usepackage{epsfig}
\begin{document}
\title{Maximisation Principles and Daisyworld}
\author{G.J. Ackland}
\affiliation{School of Physics, The University of Edinburgh, Mayfield Road, 
Edinburgh EH9 3JZ, UK 
 E-mail: gjackland@ed.ac.uk}

\begin{abstract}
{\bf We investigate whether the equilibrium time averaged state of 
a self-organising system with many internal
degrees of freedom, 2D- daisyworld, can be described by 
optimising a single quantity.
Unlike physical systems where a principle of maximum energy production  
has been observed, Daisyworld follows evolutionary dynamics rather
than Hamiltonian dynamics.  We find that this is sufficient to
invalidate the MEP principle, finding instead a different principle, 
that the system self-organises to a state which maximises the amount of life.}

\end{abstract}

\date{\today}
\maketitle

\section{Introduction}

Thermodynamics provides an excellent macroscopic description of the 
physics of matter in terms of time-independent 
averaged quantities such as energy, entropy, free energy.
These arise from the underlying microscopic, time-dependent motions of 
atoms. The two descriptions are unified through statistical 
mechanics, and therefore contain the same information, but in 
in the absence of atomic-level detail 
the thermodynamic picture is far more useful for everyday application.

It is interesting to consider whether ecosystems can similarly be described
by macroscopic averages rather than details of individual species.

Recent work has suggested that the principle of
maximum entropy production (MEP) may be a good way to investigate
complex, self organising systems.  Lorenz et al (2001,2002) 
suggest that the atmosphere of Earth and Titan may be in such a state.  
Dewar (2003) has shown
that other known distributions of open, dynamical systems, such as
self-organised criticality (Bak et al, 1988), the fluctuation theorem (Crooks, 1999) and Zipf's Law (Zipf 1932).
The basis of Dewar's argument is that if one takes an appropriate
average over all its internal possibilities\footnote{This is implicit 
in using the sum over all paths to build a partition function}, 
consistent with the known
fluxes and internal energies, MEP will emerge.  Of course, real
systems do not actually sample all possibilities, so this only
provides a sufficient conditions for MEP, the necessary condition is
that the average over possibilities actually observed converges
rapidly. This parallels an argument in equilibrium statistical
mechanics, where thermodynamic averages are calculated from an ergodic
hypothesis (that all microstates contribute according to  Boltzmann
statistics) to which the chaotic trajectories of real systems
rapidly converge.  

Boltzmann statistics follow from Hamiltonian
dynamics. In the case of ecological systems, 
the dynamics are driven by evolution, and no proof of 
convergence of an evolving
system to a Boltzmann-type distribution yet exists. It is interesting
to ask whether more general principles, such as MEP, can be applied in
an ecological context.  In this paper we explore a particular
ecological model, 2D-Daisyworld, in which the MEP state can be
rigorously defined, to see whether the system does, in fact,
self-organise thereto.  

The 
2D-Daisyworld, is based on
previous work (Von Bloh et al., 1997,1999, Ackland et al 2003).  Life is
reduced to a single species 'daisies' with a single property
(albedo) upon which selection can act. The local environment is
reduced to a single variable temperature. The planet is warmed by
a Sun whose heat production (insolation) increases slowly. Growth is
only possible across a narrow range of temperatures (5-40$^o$C) with
the mid point 22.5$^o$C being optimal and the growth rate dropping
parabolically to zero at the extremes. Local temperature is partly
determined by insolation and albedo and partly by diffusive heat flow
across the planet. Growth occurs by seeding of bare ground adjacent to
existing daisy populations, eliminating the need in the original,
zero-dimensional daisyworld of Watson and Lovelock, (1983) to predefine
different local environments for different albedo types. Each
(asexual) offspring mutates in albedo at random. 

The original daisyworld  is deterministic, cast in
the form of differential equations, and has spawned many variants
to incorporate specific evolutionary dynamics or natural selection
(Lenton, 1998, Lenton and Lovelock 2000, Lovelock 1998, Robertson 
and Robinson, 1998, Staley 2002, Sugimoto, 2002).  
The 2D version is stochastic, with
many internal degrees of freedom, allowing an entropy production to
be defined by the range of albedo (called biodiversity), as well as 
through heat flow. 

Historically, daisyworlds were designed to illustrate the 
self-regulating effects of coupling between life and the 
environment, and their impact on evolution by group selection.
This is not the issue of interest here - daisyworld is used as 
a test case for MEP.

\section{Methods} 
The temperature field changes as: 

\begin{eqnarray} C\dot{T}(x,y,t) &=& D_T\nabla^2{T}(x,y,t) \nonumber \\ 
                                 & & -\sigma_B{T}^4(x,y,t) \nonumber \\
                                 & & + S_L(1-A(x,y,t)) \end{eqnarray}

 Where $C=2500$ is the heat
capacity, T the temperature field, $D_T=D/C$ the thermal diffusion 
constant $\sigma_B=5.67\times 10^{-8}$
the Stefan-Boltzmann constant, $S_L=S/432.3$, the insolation 
normalised to that which gives T=295.5K on a bare planet
and $A$ the albedo field (daisy/bare ground). Each lattice point has an
associated temperature and albedo. $A$ ranges from 0 to 1, with 0.5
being the value for bare ground.  The daisy field is discrete and
evolves stochastically. We use a square lattice with eight
neighbouring sites. At each time step each site is examined and: 

(1) If occupied by a daisy, it changes to bare ground with probability
$\gamma(T)$
(the "death rate"). 

(2) If
occupied by bare ground, the site is populated with probability $\beta(T)$ 
(the "growth rate"). 

The new daisy 
has the albedo of its parent, from a randomly chosen neighbouring site, 
with a random fluctuation
(mutation) drawn from a uniform distribution between r and -r.  If the
neighbouring site is bare, no growth occurs.

\section{MEP or Gaia?} 

Two global principles will be advanced, these are not new, but formal 
definitions for the present system are made.

1/  Maximal Entropy Production (Dewar, 2003) - the system self organises 
to produce entropy as rapidly as possible

2/  The Gaia principle (Lovelock, 1998) - the system self organises 
to maximise the amount of life

In previous work on the 2D case (Ackland et al 2003), it was shown that 
multiple solutions satisfied the Gaia principle.  A secondary
criterion was found, maximisation of biodiversity 
entropy\footnote{Finite diffusivity leads to clump formation, which confounds this
result due to neglect the entropy associated with variation in clump
size. In principle this could be incorporated, but for the present
purposes the maximum entropy principle observed in the limit of
infinite diffusivity is sufficient.}
  $N_A \ln N_A$
where $N_A$ is the number of daisies with albedo A, leading to a mean
distribution 

\begin{equation} \langle N_A \rangle = e^{\beta A} \end{equation}

with $\beta$  given by the solution of:

\begin{equation} \langle  A \rangle \int^1_0 \exp(\beta A)dA =  
\int^1_0 A \exp(\beta A)dA\end{equation}

 Thus the global state of 2D daisyworld was classified 
in terms of two quantities, maximised
growth (i.e. temperature regulation at the favoured growth
temperature) and maximised biodiversity.  These are equivalent 
to satisfying both MEP and Gaia principles simultaneously.

\begin{figure}[htb]
\protect{\includegraphics[width=\columnwidth]{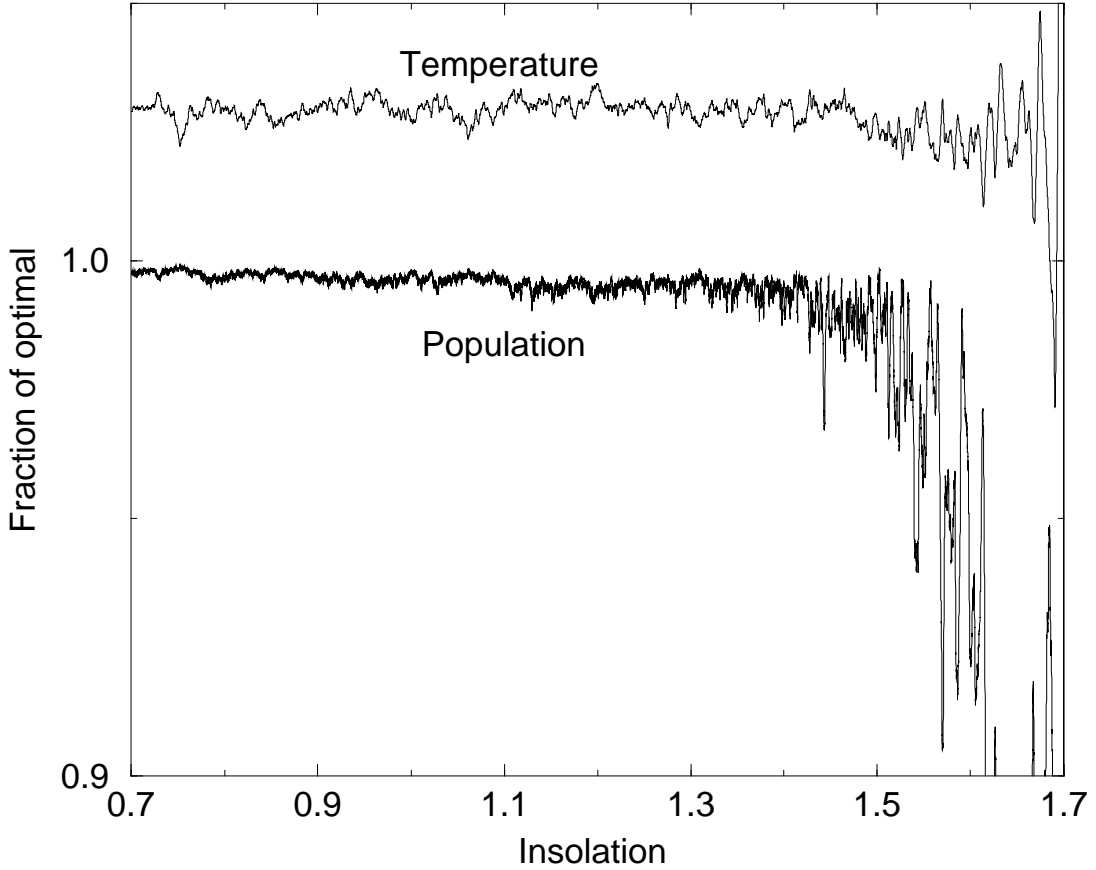}}
        \caption{\label{fig:2d2} 
Fractional daisy coverage and temperature 
(in units of optimal growth temperature,  $T_g=295.5$ ) from a calculation
with a 200x200 lattice 
($\gamma=0.02(1 - (\delta+T_d-T)(T+\delta-T_d)/2\delta^2$,
$\beta(T) =(\delta+T_g-T)(T+\delta-T_g)/\delta^2$. 
  $T_d=305.5$, $\delta=17.5) D=0.1, r=0.02$) 
with scaled insolation S varied from 0.7 to 1.7 
over $10^6$ updates of daisy and temperature fields with a temperature 
dependent death rate, minimal at $T_d=305.5$.   The graph shows that 
in the non-desert regime population tends to its maximal value 
($1-x=\gamma/(\gamma+\beta)=0.99$ 
in the mean field approximation, where regrowth is always possible) 
while temperature is regulated at $T_d$, rather than $T_g$ which would maximise(bio-)entropy production.  Thus the governing global principal appears to be Gaian 
(maximal life) rather then MEP. At higher S, fluctuations lead to long 
lived deserts, which cause the mean field approximation to break down 
as daisies cannot grow in the interior. 
}
\end{figure}

It is possible to make a small change to the 2D daisyworld so that 
Gaia and MEP cannot be satisfied simultaneously.
 To do this simply requires making the
death rate temperature dependent, and the temperature for minimum
death rate ($T_d$) different from that for maximum birthrate ($T_g$).  In
this case the MEP will predict that the temperature will adjust close
to $T_g$, while the Gaia hypothesis predicts it will adjust close to $T_d$.
\footnote{We say "close to": these are the limiting cases for low death rates
and high birth rates, assuming uncorrelated dead patches.}
  The
explanation for this is straightforward: MEP predicts that entropy
(biodiversity) be produced at the fastest possible rate: maximising
the birthrate.  By contrast, the Gaia hypothesis
is that the total number of daisies be maximised: minimising deathrate.  
Thus investigating whether the mean
temperature tends to $T_g$ or $T_d$ determines which (if either) 
of the two hypotheses
applies.  Figure.\ref{fig:2d2}  shows the variation of
mean temperature and population as a function of insolation: the mean
temperature obtained is $T_d$ and the population is $1-\gamma(T_d)$.  
Thus we conclude that the dynamics satisfy Gaia rather than MEP.

\section{Thermal Entropy Production}

The arguments above pertain to biodiversity entropy.  With a finite
diffusivity, one can also investigate whether the MEP can be applied
to entropy production by heatflow in daisyworld.  There are conceptual
difficulties with defining the system here - the models for MEP on
Earth (Ozawa et al, 2003), and Dewar's (2002)
formulation assume constant flows of
energy into and out of the system; the self-organising albedo of
daisyworld is able to regulate these flows.  Similarly, the diffusion
constant is fixed in 2D-daisyworld, whereas in atmospheric models it
is assumed to self adjust.  Notwithstanding this, there is a range of
temperatures in 2D daisyworld, and heat is transported from hotter to
colder regions, so we can define thermal entropy production\footnote{factor of two arises from double counting} as

\begin{equation}
dS/dT = \int \frac{dQ}{T} = \sum_{i} \sum^8_{k=1} \frac{D_T (T_i - T_k)}{2T_i}
\end{equation}
 
\begin{figure}[htb]
\protect{\includegraphics[width=\columnwidth]{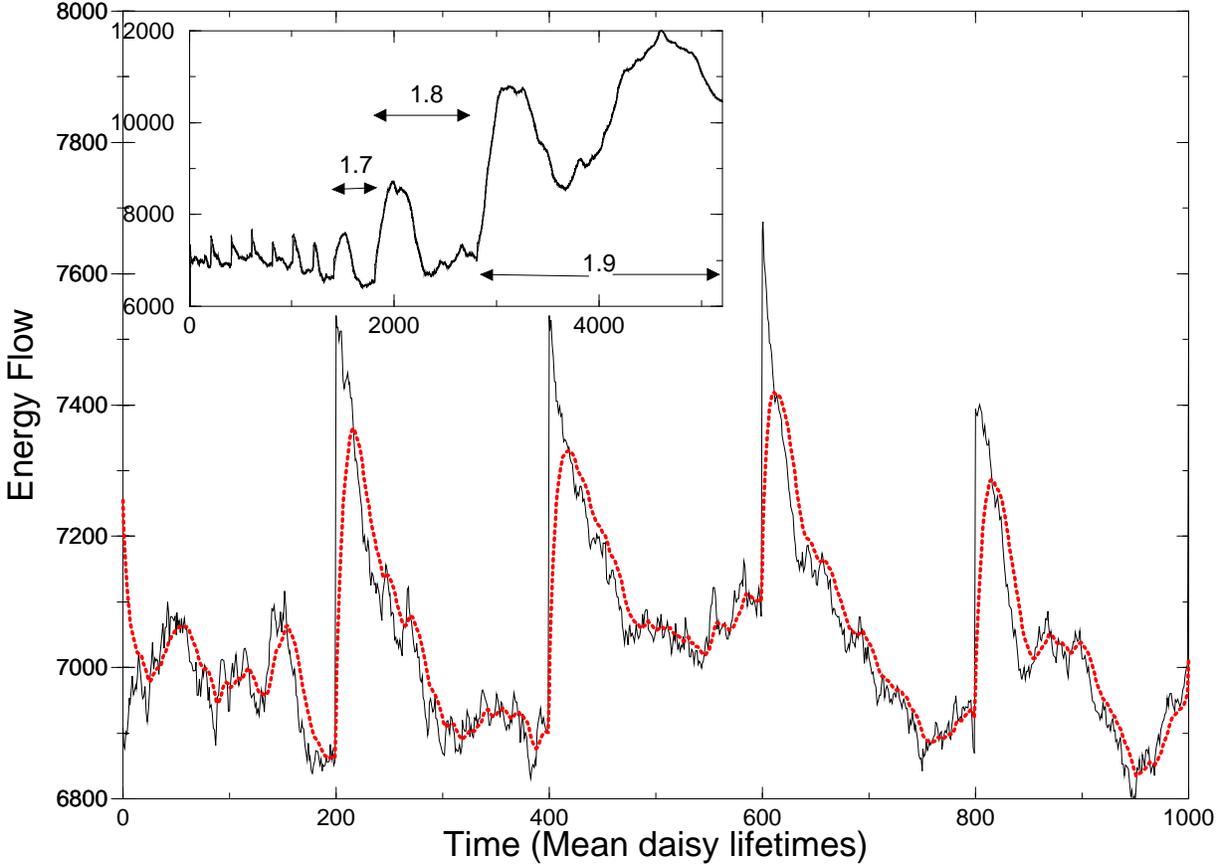}}
        \caption{\label{fig:ebalance} Graph showing net total energy flows into (solid line) and out of 
(dotted line) a 200x200 cell 
2D daisyworld system (diffusion = 0.1, death = 0.02, mutation = 0.02). 
The simulation starts with S=1.0 and increases by discrete jumps of 0.1
every 10000 timesteps.  Up to S=1.7 this is sufficient to allow the 
system to return to equilibrium: the excess 
incoming flow of energy is reduced by evolution of the albedo.  
Beyond 1.7 (see inset) the sudden change generates sufficient die back 
to create deserts in the system, which initially cause positive feedback 
and increase the incoming flow still further.  Ultimately, evolution of 
paler daisies in the non-desert regions allow reinvasion of the deserts, 
but the simulation had to be run for 20000 timesteps (S=1.7) and 50000 
(S=1.8) to achieve this.  Even in equilibrium, transient desert regions 
occur.  At S=1.9 the simulation was run for 120000 
timesteps, throughout which desert and populated regions coexisted and 
migrated around the planet (there being no curvature to favour one 
region over another).
Throughout, note that the energy emitted is smoother than that absorbed
and lags the energy absorbed by about 500 timesteps (10 mean daisy lifetimes)
$\gamma = 0.02, D = 0.1, r=0.02$.
}
\end{figure}

\begin{figure}[htb]
\protect{\includegraphics[width=\columnwidth]{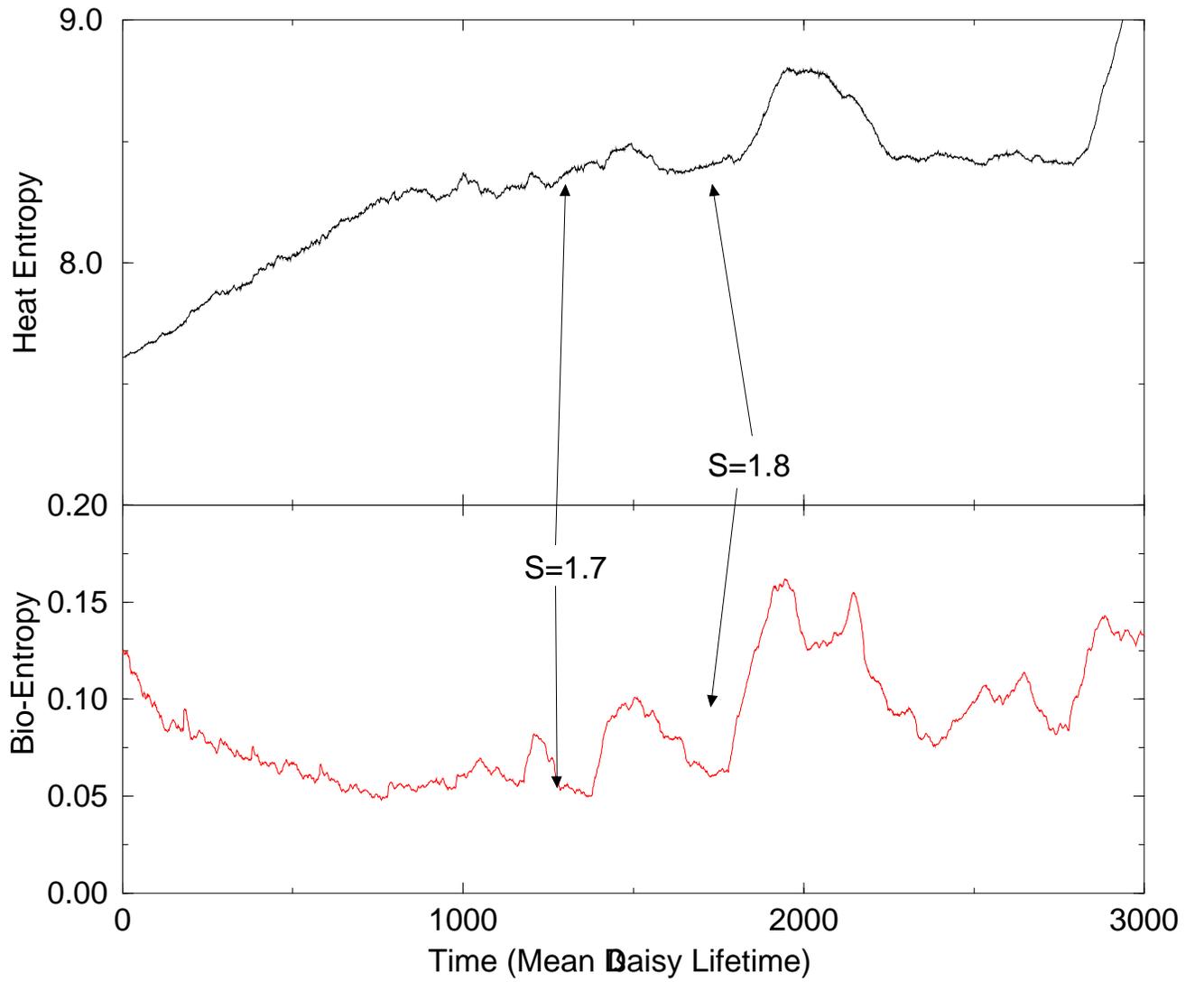}}
        \caption{\label{fig:entropy} Graphs showing variation of 
configuration (biodiversity) entropy 
 $ \langle \sum_{i=1}^{20} p(A_i) \ln p(A_i) \rangle $ from partitioning
the albedos into 20 groups, and (heat) entropy 
production $ \langle \int dQ/T \rangle  $
averaged over the 200x200 lattice and over a period of 50 timesteps
(one mean daisy lifetime) for the same calculation as Fig.\protect{\ref{fig:ebalance}}.
Each changes monotonically as the temperature is increased through the
region of full coverage.  Beyond S=1.7, where fluctuating deserts appear,
each measure becomes less well defined.  There is significantly increased 
heat flow from the hot deserts to the covered regions, so the heat flow 
becomes dominated by the area of desert.  Conversely, the biodiversity 
is reduced when much of the planet is unpopulated.
Reversing the change in temperature shows that the entropy change 
is reversible.
}
\end{figure}

Given the constraints on flow, it is not straightforward to determine
the maximum possible entropy production.  However, after
an abrupt change in the external forcing (insolation) the system will
be out of equilibrium, and will then evolve towards a steady state. If
MEP holds, the state immediately after the change should have lower
entropy production than the steady state.  Figure \ref{fig:entropy}
 shows no sign of
such behaviour.  Monitoring entropy production also enables us to
monitor another property, the net heat flow in and out of the system.
In steady state, the mean values of these heat flows should be equal 
- in 0D daisyworlds solutions initially assumed 
input and output were in balance at all times -
this is not required for a nonlinear system, but proof of the
stability of these solutions came later (Saunders, 1994).  In the
2D-daisyworld Figure \ref{fig:ebalance}
shows the time variation of flow in and flow out.  An
interesting feature here is that flow in (basically albedo) 
fluctuates much faster than flow out
(basically temperature), which also has a delayed response 
of many daisy lifetimes. 

 Observations of planetary MEP are
presumed to indicate that large-scale atmospheric heat transfer mechanisms are
self-organising.  However, even in the curved-planet
version of 2D-daisyworld (Ackland et al, 2003), 
very strong temperature regulation
eliminates large scale temperature variation and hence the driving
force for the atmosphere (except in the case of desert formation).
Figure \ref{fig:entropy}
shows that the mean heat entropy produced per cell does not tend 
to a maximum during equilibration after a sharp change in insolation.
We this conclude that no MEP principle applies to heatflow in 2D daisyworld.

We examined whether there was an optimum value for
the diffusion constant D: at given S, heat entropy
is largely independent of D.  This is non-trivial, since it
implies that for high D temperature variations are suppressed
precisely enough to compensate for the increased diffusion constant.
However, it indicates that even if D could be optimised by 
a hypothetical daisyworld atmosphere, in the presence of strong temperature
regulation no detectable maximum entropy producing value
of D exists. 

\section{ MEP in Original daisyworld }

In Watson and Lovelock's (1983) 0D
daisyworld two daisy albedos exist, denoted by suffices b and
w, with albedo $A_b=0.25$ and $A_w=0.75$ 
respectively. The relevant equations are:

Change of fractional areas occupied by each daisy type (set to zero at
equilibrium) 

\begin{equation} da_i/dt = a_i(x\beta_i-\gamma);
 \label{D1} \end{equation}

 Fraction of free surface for growth 

\begin{equation}x=1-\sum_i a_i; \label{D2} \end{equation} 

Growth rates for daisies.

\begin{equation}\beta_i={\rm max}\{0,1-((22.5-T_i)/17.5)^2 \};
 \label{D3}\end{equation} 

Mean planetary albedo (reflectivity) 

\begin{equation} A_p = x/2 + \sum_i A_ia_i; \label{D4} \end{equation} 

Local temperatures over daisyfields

\begin{equation} \label{D5}
(T_i+273)^4 = q(A_p-A_i)+(T_e+273)^4 \end{equation}

One further equation closes the set - this represents the equality of
absorbed and emitted heat, 
As we
saw in the 2D case, this equality strictly should be considered as a
time average, however if we assume that no variable large heat
reservoirs (e.g. icecaps) exist the timescales are such that the
following equality holds: 

  \begin{equation}(T_e +273)^4 = LS/\sigma_{B}(1-A_p) \label{D6} \end{equation} 

S being an insolation, L a variable of order unity, and
emitted by Stefan-Boltzmann radiation.  

\begin{figure}[htb]
\protect{\includegraphics[width=90mm]{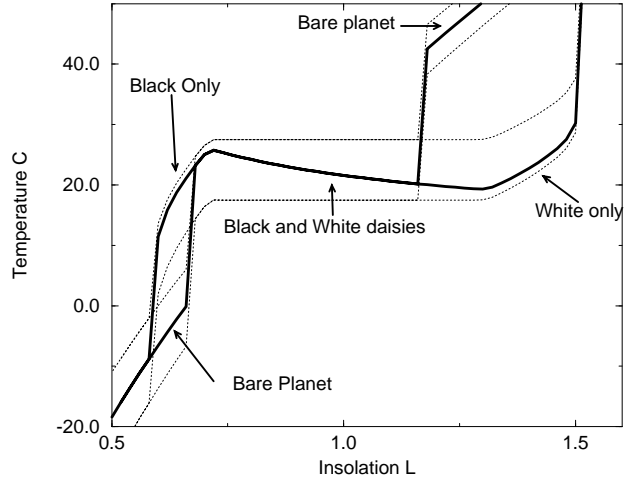}}
\newline
\protect{\includegraphics[width=90mm]{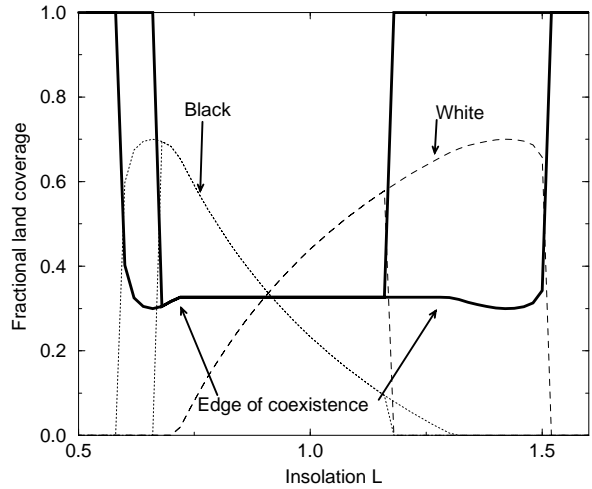}}
        \caption{\label{fig:0d2} 
Results from 0D daisyworld with parameter values from Lenton and Lovelock:
      $S/\sigma_B = 1.68\times 10^{10}K^4$,
      $q = 2.06425\times10^9$ and 
      $\gamma = 0.3$.
a) Temperature variation with increasing solar insolation.
Thick line show planetary average, dotted lines show $T_b$ and $T_w$.  
The four regimes are shown in their region of stability: Simulations
commencing with temperatures outside the growing range can stabilise
the bare planet solution, but single-daisy and two-daisy solutions do
not coexist. Only stable solutions are shown, including the bistable 
living/dead regime. 
b) coverage: Thick line denotes bare ground, thin lines denote individual species.
%
For non-zero $a_b$ and $a_w$,
$da_i/dt=0$  means that $\beta_b=\beta_w$ and 
from \ref{D3} and \ref{D5}, $T_w$ and $T_b$ are also independent of
$L$: $T_w = 22.5$ for $q=0$ falling almost linearly to zero for
$q=9.34\times10^9$. For standard parameter values from Watson and Lovelock,
$\beta_b=\beta_w=0.918$ $T_w$=17.5 and $T_b$=27.5.  
With $T_w$ and $\beta_w$ fixed by the choice of $q$,
independent of $L$, as is the amount of bare ground at equilibrium:
$x=\gamma/\beta_w$; (from 1).  Equations \ref{D2} and \ref{D4} 
now give us simple
linear expressions for $a_b$ and $a_w$ in terms of $x$ and the single
variable $A_p$.  Thus, once the assumption of species coexistence is
made, the only $L$-dependent variables are $T_e$, $A_p$ which can be
solved for analytically.  There are three unstable equilibrium solutions, 
either without daisies or with a single species Saunders (1994).
}
\end{figure}

The usual conclusion from this model is that the daisies
moderate the temperature, however 
the graph of T $vs$ L (figure \ref{fig:0d2}a): shows that 
$T_e$ is not optimised to the optimum for growth.   
In contrast, a plot of x $vs$ L  (figure \ref{fig:0d2}b) 
shows that x is a minimum for all L.  

The daisyworld equations can be solved analytically.  The solutions 
fall into four
independent classes: no daisies, black-only, white-only, both types.
At a given
insolation 
only one ``living'' solution is stable if the simulation is 
started with non-zero concentration of both daisy types.  Dead and living 
solutions coexist in regions where the dead planet temperature lies 
outside the limits for growth (5-40$^o$C).
The stable solutions (Fig.\ref{fig:0d2}) have been found dynamically, 
(iterating equations \ref{D1}-\ref{D6} to self consistency) by
numerous authors.  The important point here is that
the {\it stable} solution 
can  be determined by minimisation of $x$ among allowed 
solutions.  Adopting the lowest $x$ is the  
0D equivalent to the Gaian principle of maximum life, 
as found in the 2D case.
 
\begin{figure}[htb]
\protect{\includegraphics[width=100mm]{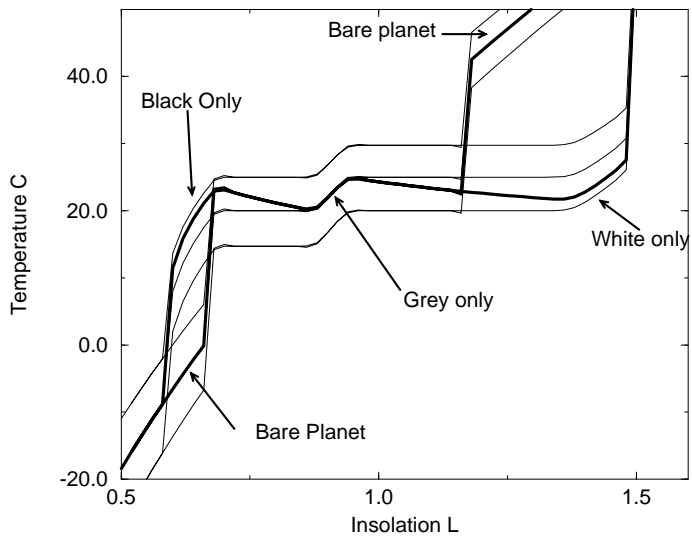}}
\newline
\protect{\includegraphics[width=100mm]{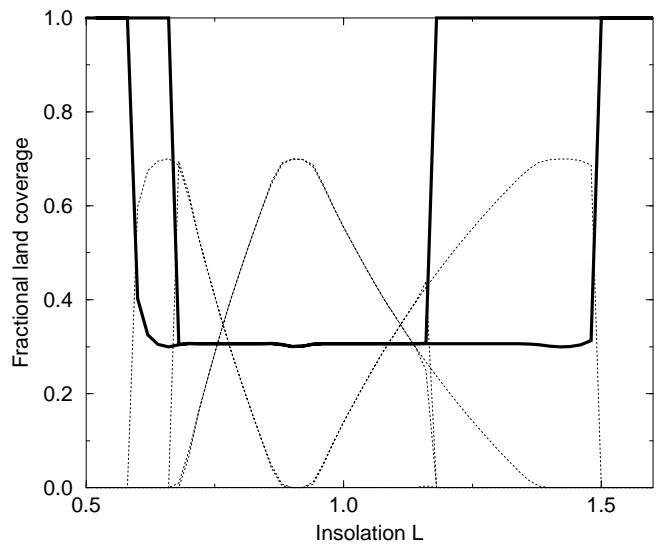}}
        \caption{\label{fig:0d3}
Daisyworld with additional grey daisy species with albedo 0.5. 
Other parameters as for Fig 4. 
a) Temperature - thick line shows planetary average, thin lines denote 
local temperatures over daisies.  Note that the grey-only solution 
is stable because it maximises life, despite having no temperature 
regulatory effect.
b) Coverage: Thick line denotes bare ground, thin lines denote 
individual species.  Note that graph is continuous, with bare ground 
(x) falling below 0.30625 only in the single-species regions.
}\end{figure}

As a further test, we introduce a third daisy species (grey) with 
albedo 0.5, the same as bare ground and therefore with no regulating effect.  
Equations \ref{D1} show that coexistence of three species is impossible, 
and in fact the stable solutions (Fig.\ref{fig:0d3}) 
traverse regions of black-only, 
black and grey, grey only, grey and white, white only.  In each region 
unstable equilibria exist for other combinations, and in each case the 
unstable equilibria have higher $x$ than the total one.  This result is 
robust against changes to  $q$, $A_i$ and $\gamma$ (including species 
dependent $\gamma$).

The equilibrium solution to daisyworld can also be obtained by 
either integrating equation \ref{D1} or with a minimisation of 
$x$ with respect to $a_i$ subject to constraints.
Each gives the same solution (Figs.\ref{fig:0d2},\ref{fig:0d3}), 
independent of the history of the system.  Thus {\it the equilibrium state 
in 2D and 0D daisyworld can be determined assuming the Gaia principle, 
without knowing 
detailed dynamics, just as the equilibrium states of thermodynamic systems 
can be determined by minimising free energy.}

\section{Gaia and MEP in the logistic Map}

The logistic map is another widely used model for population growth
(Hassell, Lawton and May, 1976).  
The resulting populations famously follow the period doubling route 
to chaos.  There is no natural selection and again it is interesting 
to ask whether this model displays any kind of maximum population/maximum 
entropy behaviour.   

The classical logistic map gives an nth population size
of

\begin{equation} 
  y(n)=ay(n-1)[1-y(n-1)]
\end{equation}

\begin{figure}[htb]
\includegraphics[width=0.9\columnwidth]{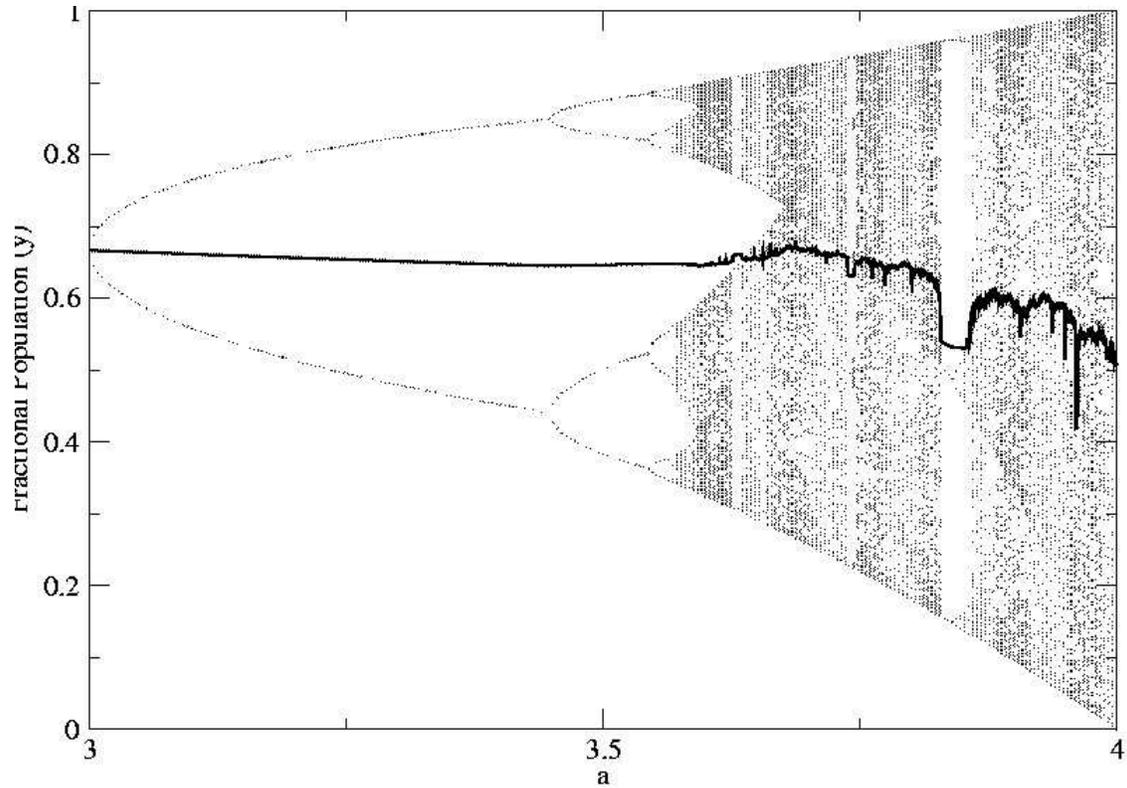}
 \caption{\label{log1}
Bold line shows the mean amount of life from the logistic map, 
plotted against recursion parameter $a$.  Background dots show the 
bifurcation diagram.  
A small amount of random noise added to $y(n)$ (drawn from a flat 
distribution $\pm 0.001$) eliminates most of the  
ordered behaviour, and the  gives mean population above 0.5 throughout.
Without noise, the mean population is still above 0.5 for all the 
chaotic cases, and for all the ordered windows except for the tiny
region between 3.9602 and 3.9616, 0.04\% of the total.  
Populations above 0.5 are found for
stable attractors across the family of logistic maps of the form 
$y_{n+1}=a[y_n(1-y_n)]^\nu$
}
\end{figure}

The equilibrium distribution for $y$  depends only on the parameter $a$. 
The mean population, averaged over 40000 iterations, is shown 
as a function of $a$ in fig\ref{log1}. It exhibits a lot of 
structure which can be connected with the bifurcation diagram of the 
logistic map.  The mean population of the attractor is higher than would be 
expected from maximum entropy (all levels equally occupied). 

Turning now to entropy, there are many definitions related to  the 
population distribution by analogy with the 2D daisyworld.  For example, 
the Tsallis entropy (Latora et al 2000)

\begin{equation}
S = \sum_{i=1}^N log(P(y_i)) / Nlog(M/N)
\end{equation} 

where $P(y_i)$ is the number of occurences of a population 
between $\frac{i-1}{N}$ and $\frac{i}{N}$ during M iterations 
of the logistic map (taken after a 1000 iteration equilibration period).

This quantity has a maximum possible value of 1, and 
(neglecting numerous distinct special cases of 
limit cycles) its value tends relatively smoothly to this value 
with increasing $a$. However, its actual 
value depends on N.  This only 
tells us that the area lying within the attractor increases with $a$.
  
The attractors in the logistic map provide some constraints 
on the mean value of $y$ and the Tsallis entropy: it is 
difficult to define what the 
``maximum life'' solution would be.  Thus while it is notable that 
perturbed logistic map dynamics gives mean population in its 
stable region above 0.5 for all $a$, no quantitative principle emerges.

\section{Conclusions} The 2D-daisyworld model has been used as a
testbed to determine whether any global maximisation is applicable to a
system governed by the dynamics of natural selection.  
We find that Maximum Entropy Production is not valid here, either 
for biodiversity or heat transfer entropy.  This casts doubt on some
recent work in which the MEP is assumed as a constraint and used to
obtain a solution (e.g. Pujol, 2003).  Analysis of 2D and 0D cases 
suggests that daisyworlds appear to obey a separate 
optimisation principle, in the simple 2D case indistinguishable from 
MEP, that of maximising the amount of life present
in the system.  The logistic map attractors have above average population, 
but for this system it is difficult to define what the maximal population 
value should be.

The more fundamental implication is that stability of a
dynamic equilibrium can be determined as an optimisation problem on
the properties of the system as observed, and that detailed
integration of system dynamics through time may be unnecessary in
determining the equilibrium state.

\section{References}

P. Bak, C. Tang, and K. Wiesenfeld, "Self-organized
criticality", Physical Review A 38 (1988) 364

 G. E. Crooks, 
 "The Entropy Production Fluctuation Theorem and the Nonequilibrium Work Relation for Free Energy Differences",
 Phys. Rev. E 60, 2721 (1999),

 Dewar, R.C. 
Information theoretic explanation of maximum entropy production, the fluctuation theorem and self-organized criticality in non-equilibrium stationary states.
J.Phys.A 36 631 (2003)


Hassell, M. P., Lawton, J. H. and R. M. May. 1976. Patterns of dynamical behavior in single species populations. J. Animal Ecology. 45: 471-476.

Kirchner, J.W. (2002). The Gaia hypothesis: Fact, theory, and wishful thinking. Climatic 
Change 52, 391-408.

Lenton, T.M. (1998). Gaia and natural selection. Nature 394, 439-447.

Lenton, T.M. and Lovelock, J.E. (2000). Daisyworld is Darwinian: Constraints on 
Adaptation are Important for Planetary Self-Regulation. Journal of Theoretical Biology 
206, 109-114.


 Lorenz, R.D. (2002)
Planets, life and the production of entropy.
International Journal of Astrobiology 1, 3-13.

Lorenz, R.D., Lunine, J.I., Withers, P.G. and McKay, C.P. (2001)
Titan, Mars and Earth: Entropy Production by Latitudinal Heat Transport.
Geophys. Res. Lett. 28, 415-418. 

Lovelock, J.E. (1988). The Ages of Gaia - A Biography of Our Living Earth. New 
York: W. W. Norton and Co.


Latora,V.  Baranger,M.  Rapisarda,A. and C. Tsallis,C. "Generalization to nonextensive systems of the rate of entropy increase: the case of the logistic map", Physics Letters A 273 (2000) 97. 

May, R. M. 1976. Simple mathematical models with very complicated dynamics. Nature. 261: 459-467.

Ozawa,H. Ohmura,A. Lorenz,R.D. Pujol,T.
The second law of thermodynamics and the global climate system - A review of maximum entropy production principle. Reviews of Geophysics (accepted)

Prasad, A, Mehra, V and Ramaswamy, R.  Strange nonchaotic attractors 
in the quasiperiodically forced logistic map
Physical Review E, 57, 1576-84 (1998)

 Pujol, T. (2002)
The consequence of maximum thermodynamic efficiency in Daisyworld.
J. theor. Biol. 217, 53-60. 


Robertson, D. and Robinson, J. (1998). Darwinian Daisyworld. Journal of Theoretical 
Biology 195, 129-134.

Saunders, P.T. (1994). Evolution without Natural Selection: Further Implications of the 
Daisyworld Parable. Journal of Theoretical Biology 166, 365-373.

Staley M
    Darwinian selection leads to Gaia in the long run
J Theor Biology 218 35 2002 

Sugimoto, T. (2002). Darwinian evolution does not rule out Gaia hypothesis. Journal of 
Theoretical Biology.218, 447-455.

Von Bloh, W., Block, A. and Schellnhuber, H.J. (1997) Self-stabilization of the biosphere 
under global change: a tutorial geophysiological approach. Tellus 49B, 249-262.

Von Bloh, W., Block, A. and Schellnhuber, H.J. (1999) Tutorial Modelling of geosphere-
biosphere interactions: the effect of percolation-type habitat fragmentation. Physica A 
266, 186-196.

Watson, A.J. and Lovelock, J.E. (1983). Biological homeostasis of the global 
environment: the parable of Daisyworld. Tellus 35B, 284-289.

Zipf, G.K  Selective Studies and the Principle of Relative Frequency in Language  (Harvard University Press, Cambridge, 1932)

\end{document}